\documentclass[aps,prd,twocolumn,groupedaddress]{revtex4}

\usepackage{graphicx}% Include figure files
\usepackage{dcolumn}% Align table columns on decimal point
\usepackage{bm}% bold math
\renewcommand\({\left(}
\renewcommand\){\right)}

\def\lsim{\raise 0.4ex\hbox{$<$}\kern -0.8em\lower 0.62
ex\hbox{$\sim$}}

\def\gsim{\raise 0.4ex\hbox{$>$}\kern -0.7em\lower 0.62
ex\hbox{$\sim$}}

\def\lbar{{\hbox{$\lambda$}\kern -0.7em\raise 0.6ex
\hbox{$-$}}}

\newcommand\eq[1]{eq.~(\ref{#1})}

\newcommand\ee{\end{equation}}
\newcommand\be{\begin{equation}}
\def\bea{\begin{eqnarray}}
\def\eea{\end{eqnarray}}
\newcommand\ees{\end{eqnarray}}
\newcommand\bees{\begin{eqnarray}}

\def\p1{{\bf p}_1}
\def\p2{{\bf p}_2}
\def\k1{{\bf k}_1}
\def\k2{{\bf k}_2}

\newcommand{\hti}{\tilde{h}}

\newcommand{\fmax}{f_{\rm max}}

\newcommand{\msun}{M_{\odot}}

\def\D{\Delta}
\def\p{\partial}

\def\nn{\nonumber}

\def\eps{\epsilon}

\def\dslash{\hspace{-1mm}\not{\hbox{\kern-2pt $\partial$}}}
\def\Dslash{\not{\hbox{\kern-4pt $D$}}}
\def\pslash{\not{\hbox{\kern-2.1pt $p$}}}
\def\kslash{\not{\hbox{\kern-2.3pt $k$}}}
\def\qslash{\not{\hbox{\kern-2.3pt $q$}}}

\begin{document}

%\preprint{UGVA-DPT 2003/xxx}

\newcommand{\hrss}{h_{\rm rss}}

\title{Upper limits on the gravitational mass loss of the Galaxy\\
and the  LIGO burst searches}

\author{Florian Dubath}
\author{Michele Maggiore}
\affiliation{D\'epartement de Physique Th\'eorique, 
Universit\'e de Gen\`eve, 24 quai Ansermet, CH-1211 Gen\`eve 4}

\date{\today}

\begin{abstract}
We discuss the relevance, for the search of gravitational-wave bursts,
of upper limits on the total mass loss of the Galaxy which come from
various astronomical observations. For sub-millisecond bursts
we obtain limits on the event rate, as a function of the GW amplitude,
which are stronger
than the corresponding
upper limits set by LIGO in the S2 run.
A detection of a burst rate saturating these limits, with the
sensitivities of present and near-future runs, 
would imply that, with some improvement on the accuracy 
of astronomical
observations of the Galaxy, as foreseen with the GAIA mission, 
it might be possible to detect gravitational waves
indirectly from their effect
on galactic dynamics.

\end{abstract}

% insert suggested PACS numbers in braces on next line
% insert suggested keywords - APS authors don't need to do this
%\keywords{}

%\maketitle must follow title, authors, abstract, \pacs, and \keywords
\maketitle

\section{Introduction}

One of the main
target of the existing gravitational-wave (GW) detectors are short bursts,
with a duration between fractions of a millisecond and a few seconds, 
which could
originate from astrophysical events. The LIGO collaboration has
published the results of the S2 run~\cite{LigoS2}, (see also 
\cite{LigoS3} for the S3 run)
which extended the search to lower
values of the GW amplitude, compared
to  previous searches by resonant-mass 
detectors \cite{ROG01,IGEC,ROG03}. 
Presently, LIGO is performing a long data-taking
run at its target sensitivity which, in case of no positive 
detection, should anyway produce a  more significant bound on the event
rate. 

The result of these searches  can be presented as an upper limit on
the event rate of GW bursts, ${\cal R}$,
as a function of the strength of the GW signal. The latter can be conveniently
characterized in 
terms of the so-called root-sum-square amplitude $\hrss$, defined by
\be
h^2_{\rm rss}\equiv \int_{-\infty}^{\infty}dt\, h^2(t) \, .
\ee
The purpose of this note is to point out that a bound on
${\cal R}$ as a function of $\hrss$ can also be obtained from considerations
of galactic dynamics, and it is in fact quite significant compared to the
bounds that can be obtained with  existing and near-future sensitivities of
GW detectors, especially for bursts of short duration, say $\tau\simeq
0.1$~ms. 

The bound emerges from the fact that, if there is a steady rate of GW bursts, 
the Galaxy has a corresponding rate of mass loss into GWs. Since it is
difficult to imagine that the burst rate today  is significantly higher than
in the past, one must consider the cumulative effect of this mass loss over a
period comparable to the age of the Galaxy, and this can have significant
consequences on various aspects of galactic dynamics. 
These issues were first addressed many
years ago~\cite{Kerr,SFR,PA,OB}, and in \cite{CDM} we reconsidered them,
using the present
knowledge of galactic dynamics, and we showed that much more stringent bounds
emerge nowadays. Let us summarize the results, referring the
reader to Ref.~\cite{CDM} for details and more references. The main 
observations that
allow us to put a bound are discussed below.

\section{Upper limits from galactic dynamics}

(i) {\em Effect of the mass loss on the radial velocity of stars.}
If the Galaxy is loosing mass, stars become less and less bound and
acquire radial velocities with respect to the Galaxy rest frame.
This modifies the radial velocity of stars, $v_r$, 
inducing a so-called $K$-term,
\be\label{ur2}\label{vr}
v_r = A R \sin 2l +K R\, ,
\ee
where the term $A R \sin 2l$ is the standard effect due to the differential
rotation of the Galaxy
($A$ is Oort's constant, $R$ is the distance of the star from the sun, and $l$
the galactic longitude), and the effect of mass loss is in the second term,
where $K=-\dot{M}/M$,
and $M$ is the mass of the Galaxy.
From the experimental bound on $K$, we
deduced in \cite{CDM} a bound
\be\label{bound1}
-\dot{M}< O(30) \msun /{\rm yr}\, .
\ee

(ii) {\em Mass loss and outward motion of the LSR}.
Rather than looking at the $K$ term, i.e. at
the expansion/contraction of the stars within a few kpc from the sun,
one can  investigate whether the local standard of rest (LSR) has an
overall outward radial velocity, as suggested by \eq{vr}.
Here, the most interesting  information comes
from the  observation of the
21 cm absorption line toward the galactic center~\cite{RS}, which shows
that the gas along the line-of-sight  has a mean 
radial velocity with respect to the LSR of  
$-0.23\pm 0.06$ km/s. The absorbing material is probably at 1-2 kpc from the
galactic center. A radial expansion 
due to mass loss predicts a radial velocity
$v_r\sim r$, see \eq{vr}, where $r$ is the radial distance from the galactic
center,  and therefore we should expect a
difference in velocity
between us and this gas, $\D v_r =(-\dot{M}/M)\D r$, where 
$\D r\simeq 6$~kpc is the distance of the Sun from the absorbing gas.

In general, there can be  both positive
and negative contributions from different physical mechanisms
to the value $\D v_r =-0.23\pm 0.06$ km/s, and to
extract a bound on mass loss to GW 
we require that no fine tuning between different contributions
takes place. We set conventionally  at 20\% the maximum fine
tuning that we allow, which means that
we say that a positive contribution from GWs
to $\D v_r$, if it exists at all, 
must be smaller than $ O(0.04)$~km/s. Setting the distance
between us and the gas to $\D r= 6$~kpc, this gives a bound 
\be\label{005}
-\dot{M}< O(0.5)\,  \msun/{\rm yr}\, .
\ee
Of course, precise numbers depend on the level of fine tuning that we 
can tolerate,
but it is clear that we cannot raise this bound by, say, one order of
magnitude, without invoking very unnatural cancellations between completely
unrelated phenomena.

(iii) {\em Upper limits from globular clusters}.
Similar bounds  have been found
using globular clusters as probes~\cite{PA,DR}. The
idea is that, if the mass of the Galaxy was much bigger in
the past, the orbits of globular clusters would have been much closer
to the galactic nucleus, and this close interaction with a very
massive central nucleus would have produced the tidal disruption of
the  cluster. The analysis of five different globular clusters
gives the result~\cite{DR}
\be\label{OmCen}
-\dot{M}< O(10)\,  \msun/{\rm yr}\, .
\ee

(iv) {\em Old wide binaries}.
Another limit comes  from the existence of old wide binaries, since for
a very massive galactic nucleus  the galactic orbits would have
been much smaller than at present. Therefore the density of stars
would have been much larger and the dissolution time of binaries due
to stellar encounters correspondingly shorter. From a list of 11 
well observed old wide binaries one finds a
limit on steady mass loss~\cite{PA}
\be
-\dot{M}< O(10)\,  \msun/{\rm yr}\, .
\ee
In conclusion, we have four different methods which all give a bound
on $-\dot{M}$ between $O(1)$ and $O(10) \msun$, so we write
\be\label{1dotM}
-\dot{M}< \eps\,\,  \msun/{\rm yr}\, ,
\ee
and we expect $\eps \sim 1$, unless one cannot find a way, perhaps with some
fine tuning, to relax 
the most stringent bound (\ref{005}).
Anyway, given that we have three more independent limits on
$\eps$, we see that we cannot stretch the value of $\eps$ beyond, say,
$\eps\simeq 10$. Furthermore, independently of the technical details leading
to the above bounds, it is easy to understand qualitatively why a bound 
on $-\dot{M}$ of
order $1\msun$/yr emerges. The total mass  of the galactic
disk plus bulge and spheroid is estimated to be $9\times 10^{10}\msun$, while
a lower bound on the age of the Galaxy is provided by the age of its
oldest globular clusters, which is $1.2\times 10^{10}$~yr. Therefore
a mass loss rate 
due to GWs of the
order of a few solar masses per year implies a total mass loss, over the age of
the Galaxy, comparable to its present mass. It is not surprising that, at this
level, one finds important dynamical effects related  mass loss. 
If one streched $\eps$ up to values of order 10, one should admit that 
the mass that the galactic disk 
has lost to GWs during its history is larger than its
present mass, so over 50\% of the original mass would have been lost to GWs.

We  also remark that these are just upper bounds and, as far as the
above arguments are concerned, there is no
physical reason that suggests that they might be almost saturated, so they
should not be taken as an indication of a plausible value of 
$\dot{M}$.

\section{Implications for galactic  bursts of GWs}

\begin{figure}
\includegraphics[width=0.4\textwidth]{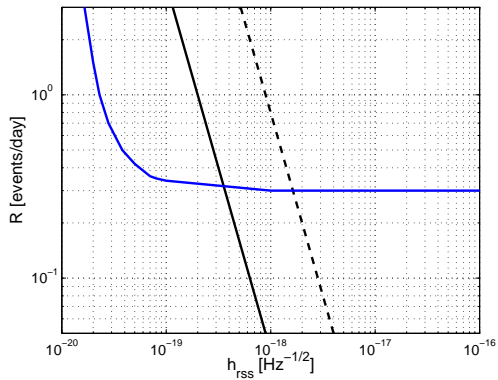}
\caption{\label{fig1} The limit from galactic dinamics with $\eps =0.5$
(solid line) and $\eps =10$ (dashed), compared to the bound from the LIGO S2 
run, when $\tau =1$~ms. LIGO data taken from Fig.~12 of Ref.~\cite{LigoS2}. }
\end{figure}

\begin{figure}[t]
\includegraphics[width=0.4\textwidth]{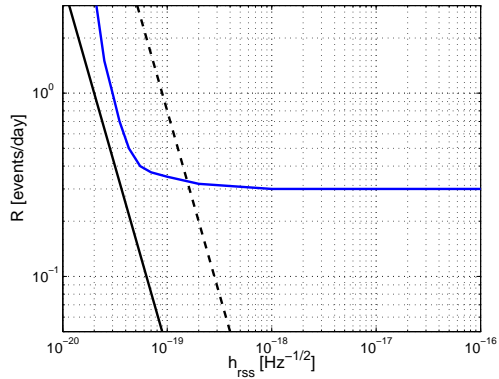}
\caption{\label{fig2} The same as Fig.~1, with $\tau =0.1$~ms. }
\end{figure}

We now compare this bound, which is generic and holds  whatever is the physical
origin of the mass loss of the Galaxy, with the energy carried away by GWs, if
there are GW bursts with a typical rss amplitude $\hrss$ 
and a rate ${\cal R}$. Following Ref.~\cite{LigoS2}, we consider a Gaussian
waveform of duration $\tau$,  given by
\be
h(t) =\hrss \(\frac{2}{\pi\tau^2}\)^{1/4}\, e^{-t^2/\tau^2}\, ,
\ee
whose Fourier transform is
\be\label{Fou}
\tilde{h}(f) =\hrss (2\pi\tau^2)^{1/4}\, \exp (-\pi^2\tau^2 f^2 )\, .
\ee
(The same  analysis can be repeated with sine-gaussian waveforms, with
similar results). We first consider a wave coming from optimal direction with
$h_+(t)=h(t)$ and $h_{\times}(t)=0$. From the standard expression of the
energy flux,
\be\label{1dEdAdf} 
\frac{dE}{dAdf}=\frac{\pi c^3}{2G} f^2 
\(  |\hti_{+}(f) |^2 +|\hti_{\times}(f) |^2  \)\, ,
\ee
where $dA =r^2d\Omega$ and $r$ is the distance to the source,
we get that the total energy radiated by such a burst,
\be
E =4\pi r^2 \int_0^{\infty} \frac{dE}{df} = 
 \frac{r^2 c^3h^2_{\rm rss}}{4G\tau^2}\, .
\ee
The average over the square of the pattern function of the interferometer gives
the usual factor 2/5 which, following the conventions of
the LIGO collaboration, we include in the definition of $\hrss$.
Then, the  energy carried by a burst, averaged over
the arrival direction and the
polarization, is
\bees
\langle E\rangle &=& \frac{r^2 c^3h^2_{\rm rss}}{4G\tau^2}
\simeq  3.4\times 10^{-4}\msun c^2\, 
\(\frac{r}{8\, {\rm kpc}}\)^2\nn\\
&&\times\(\frac{\hrss}{10^{-19}\, {\rm Hz}^{-1/2}}\)^2\,
\(\frac{1\, {\rm ms}}{\tau}\)^2\, ,\label{lEr}
\ees
where in the second line we normalized $r$ to  the distance 
to the  galactic center. We see from the above that 
a burst of duration 1~ms,
with $\hrss =10^{-19} \, {\rm Hz}^{-1/2}$, carries away about
$3\times 10^{-4}$ solar masses  in GWs, if it comes from a source located at
typical galactic distances. Of course shorter bursts are more energetic, since 
they extend in frequency space up to $\fmax\sim 1/\tau$.  In an astrophysical
context it makes sense to consider bursts as short as $\tau \simeq 0.1$~ms,
corresponding, for the gaussian wavepacket (\ref{Fou}), to $\fmax$ 
of order of a few kHz, which indeed are the shortest burst
searched for in Refs.~\cite{LigoS2,LigoS3}. If ${\cal R}$
is the rate of these bursts,
the associated
rate of mass loss of the Galaxy  is
$\dot{M} = - {\cal R} \times \langle E\rangle /c^2$, with
$\langle E\rangle$ given by \eq{lEr}. Combining this with
\eq{1dotM} we get
\be\label{1rate3}
{\cal R} <8.0\, \eps \, \frac{{\rm events}}{\rm day}\, 
\(\frac{8\, {\rm kpc}}{r}\)^2
\(\frac{10^{-19}\, {\rm Hz}^{-1/2}}{\hrss}\)^2\,
\(\frac{\tau}{1\, {\rm ms}}\)^2\, ,
\ee
where now $r$ is an average distance scale characterizing the population of
sources~\footnote{More precisely, we are  considering 
an ensemble of bursts such that the average of the square of their
rss amplitude is
$h^2_{\rm rss}$, and  $r^2$ is
the average of their squared distance. We are also
assuming an isotropic distribution of sources, since $ \langle E\rangle$ is
averaged over all directions in the sky. This assumption
should be correct
to within a factor of order unity, which  is not a significant
error for the order-or-magnitude estimates in this paper.}.  
In Fig.~\ref{fig1} we show this bound, in the 
$({\cal R} , \hrss )$ plane, for $\tau =1$~ms, 
$r=8$~kpc, and we
compare it to the experimental bounds obtained by the S2 LIGO run, 
for the same value of $\tau$.  
The solid line corresponds to  $\eps =0.5$, and the dashed line to
$\eps =10$.
In Fig.~\ref{fig2} we perform the same comparison for 
$\tau =0.1$~ms. 

We see that the bound from galactic dynamics is a significant one, especially
for small $\tau$, and indeed for $\tau =0.1$~ms is quite stronger than the 
LIGO S2 result. With the on-going and future high-sensitivity runs by LIGO and
VIRGO one can
expect that, even for $\tau =0.1$~ms, one will go beyond the small-$\hrss$ 
portion
of the limiting curve determined from
galactic dynamics, while at larger $\hrss$ the limit from galactic dynamics
will remain the dominant one.
If a statistically significant rate in excess of the value
obtained from \eq{1rate3} with $\eps \simeq 1$ and $r\simeq 8$~kpc
were found and if, from the energetic of the events, one concluded that
they originated in our Galaxy,  then one should study the possibility of
relaxing somehow the bound on $\eps$ (which anyway should be possible
at most by one order of
magnitude). Otherwise, one could consider the possibility that, rather than
having a homogeneous distribution of sources in the galactic disk, the signal
could be  due to a single (or a few) source at a close distance from us,
which emits repeatedly GW bursts~\cite{CDM,Dubath:2004vv}.
Therefore,
\eq{1rate3} can give useful clues as to the spatial distribution and possibly
the origin of the sources.

Conversely, if GW detectors should find that
the   bound that we have discussed 
is saturated,
this would mean that the effect of GW emission on some astronomical
observables,
such as the radial motion of the LSR, is just 
of order 10-20 \% of
the present observational uncertainties.
This
would imply that, with an  increase in the accuracy
of  astronomical observations, as is expected with the GAIA mission
(see e.g. Ref.~\cite{gaia}),
one could be able to single out the effect of GW
emission on the dynamics of the Galaxy, providing a form of indirect detection
of GWs.

\vspace{2mm}

Our work is partially supported by the Fond National Suisse.
We thank Erik Katsavounidis for useful comments.

\end{document}